Increasing retrofit device adoption in social housing: evidence from two field experiments in Belgium

Mona Bielig[1], Celina Kacperski[1,2], Florian Kutzner[1]

[1]Seeburg Castle University

[2]Konstanz University

Corresponding Author: Mona Bielig

e-mail: mona.bielig@uni-seeburg.at

Author Note

Financial support by the European Union Horizon 2020 research and innovation programme is gratefully acknowledged (Project DECIDE, grant N894255). We would further like to thank our colleagues at ThermoVault for supporting the design, development and implementation of the study, including their detailed knowledge and description about the technology.

Abstract

Energy efficient technologies are particularly important for social housing settings: they offer the potential to improve tenants' wellbeing through monetary savings and comfort, while reducing emissions of entire communities. Slow uptake of innovative energy technology in social housing has been associated with a lack of trust and the perceived risks of adoption. To counteract both, we designed a communication campaign for a retrofit technology for heating including social norms for technology adoption and concretely experienced benefits. We report two randomized controlled trials (RCT) in two different social housing communities in Belgium. In the first study, randomization was on housing block level: the communication led to significant higher uptake rates compared to the control group, (ß = 1.7, $p = .024$). In the second study randomization occurred on apartment level, again yielding a significant increase (ß = 1.62, $p = 0.02$), when an interaction with housing blocks was considered. We discuss challenges of conducting randomized controlled trials in social housing communities.

Increasing retrofit device adoption in social housing: evidence from two field experiments in Belgium

**Introduction**

The EU has set ambitious targets to reduce its greenhouse gas emissions and improve energy efficiency in housing, as the residential sector accounted for 27% of energy consumption in the EU in 2021 (European Environment Agency (EEA), 2021). Retrofitting existing buildings and infrastructure with energy-efficient heating technologies can significantly contribute to achieving climate goals, with space heating being the main use of energy by households (64%; (European Environment Agency (EEA), 2021). Additionally, benefits for energy justice can be expected, considering that "*it's often the most vulnerable who live in the least efficient houses and therefore struggle to pay the bills*"[1].

In reality, installation of energy-efficient technologies is often stalled by social barriers (McCabe et al., 2018). Rejection of these technologies seems particularly prominent in social housing settings (Schleich, 2019), where improvements in energy efficiency are a key priority (OECD, 2020) and where their adoption would not only increase their sustainability but also yield savings and increases in well-being (McCabe et al., 2018).

Together with a provider of a retrofit technology for space and water heating, we developed a low-cost intervention to boost retrofit technology acceptance. We employ descriptive social norms of adoption rates and experienced benefits of technology uptake (based on tenants data in the same social housing company), and investigate whether a communication intervention towards social housing tenants can increase the adoption rate of a retrofit technology in two randomized controlled trials (RCTs).

---

[1] Statement by Kadri Simson, European Commissioner for Energy, 15.12.2021

# Theoretical Background

Energy efficiency and retrofitting technologies have the potential to significantly reduce carbon emissions and generate economic savings in the social housing sector. Prior research has demonstrated the efficacy of smart heating retrofit technology (Fabrizio et al., 2017) in achieving these goals. Beyond their ecological impact, these technologies have the potential to improve the living conditions of tenants by enhancing thermal comfort, well-being, and health (Coyne et al., 2018; McCabe et al., 2018; Teli et al., 2016).

**Barriers to adoption**

However, the potential for energy efficient technology in social housing settings is often not met. One study estimated the difference between potential versus actual $CO_2$ reduction in a sustainable social housing program at around 20% less than expected – rejection by tenants was reported as the main reason (Hernandez-Roman et al., 2017). Qualitative research studies describe a lack of understanding, knowledge, and engagement among social housing tenants as well as concerns about operation for retrofit technology and renewable heating (Brown et al., 2014; Moore et al., 2015). One study (N = 251) identified concerns related to the technology itself, and the installation process, as the primary barriers to adoption (Chahal et al., 2012). A systematic review (McCabe et al., 2018) concludes that among the most relevant barriers for the uptake of renewable energy related technology in social housing are particularly lack of trust in benefits, lack of tenant engagement and perceived risk of technology adoption. In this context, successful communication for adoption of sustainable retrofit devices has to be shaped to withstand additional focus on perceived risks and tenants' technology insecurity (Brown et al., 2014; Chahal et al., 2012; McCabe et al., 2018).

**Social norms**

When mental models about behavior are characterized by uncertainty, people are more prone to look towards behaviors of others to guide their decision-making (Farrow et al., 2017; Melamed et al., 2019). Behavioral theories as the theory of planned behavior (Ajzen, 1985, 1991) or theory of normative social behavior (Real & Rimal, 2005) have long acknowledged this influence of the social environment emphasizing the role of perceived social approval and group behavior in guiding individuals' actions and choices. Social norms, defined as the understood rules and standards within a group that guide behavior (Cialdini & Trost, 1998) can be classified into in injunctive and descriptive social norms: while the former describes an individual's perception of what others expect them do, i.e. normative belief, the latter refers to an individual's belief or knowledge about the prevalence of a behavior (Cialdini, 2007; Cialdini et al., 1990; Deutsch & Gerard, 1955; Rivis & Sheeran, 2003).

**Social norm information and energy behavior**

Particularly the field of energy conservation has increasingly recognized the pivotal role of social norms as intervention for behavior change. Several field studies find that communicating descriptive and/or injunctive norms has significant impact on energy related practices in public (Bator et al., 2014; Bergquist & Nilsson, 2016; Dwyer et al., 2015; Liu et al., 2016) and residential settings (Allcott, 2011; Ayres et al., 2013; Bhanot, 2021; Schultz et al., 2007, 2015). In different meta-analyses (see Appendix B for an in-depth overview of studies), an overall positive effect of norms on energy conservation behavior and adoption of energy related technologies was found (Cialdini & Jacobson, 2021; Composto & Weber, 2022; Farrow et al., 2017). However, the evidence for effectiveness is mixed, suggesting the need for careful design, implementation and adaptation to the specific target group (Composto & Weber, 2022;

Constantino et al., 2022), taking into account moderators like social identification or proximity of behavior (Abrahamse & Steg, 2013; Cialdini & Jacobson, 2021).

**Combining social norm and benefit information**

The communication of norms is particularly relevant for the given context, as the adoption of a retrofit technology is a private behavior (Heffetz, 2012) and false assumptions about other tenants' adoption rates might further hinder adoption (Bal et al., 2021). Further, social norm information about a technology uptake should be connected to experienced benefits of others: perceived benefits of a behavior moderate the relationship between descriptive norms and behavioral intention, i.e. research found that norms did not work when individuals did not perceive the benefits of engaging in the target behavior (Rimal et al., 2005). High credibility of information source (social contacts) has further been found to interact with information on retrofit benefits to increase homeowners' cooperation (Jia et al., 2021), pointing to an interaction of social norms with benefit information.

We therefore aimed to investigate whether an intervention in which we communicate the concrete benefits through a descriptive norms of uptake and experienced benefits from other tenants can boost adoption of a retrofit technology in social housing. Critically, research targeting adoption of energy related technology in social housing is still scarce, and constrained by a case study approach (McCabe et al., 2018). Based on clear need for more experimental evidence in energy research (Sovacool et al., 2018), we designed a communication intervention for a retrofit technology rollout in social housing and tested the effectiveness of this intervention in two field studies. We hypothesized that adding a descriptive social norm information of prior technology uptake and resulting benefits by others in the social housing neighborhood will increase retrofit device uptake.

# Methodology

Both studies were conducted in collaboration with ThermoVault, a technology provider offering retrofit solutions for electrical space and water heaters to improve energy efficiency. ThermoVault collaborates with social housing companies in Belgium, enabling them savings on maintenance costs and energy savings for their tenants. Social housing tenants receive the device for free from their social housing management. Despite the advantages, a significant proportion of tenants in the past have rejected the technology, ranging from 20-30%[2]. Building on this, we aimed to create a low-cost intervention to increase the acceptance of the technology in the social housing context. For this, we leveraged the communication strategy employed by ThermoVault, where tenants receive a letter informing them of the benefits and installation date prior to technology rollout. This letter served as basis for implementing our experimental intervention. The study was carried out in line with ethics requirements of the German Ethics Board (DGPS) as well as European data protection guidelines (DGPR). Due to the nature of the field trial, it was not possible to obtain individual consent from each tenant; instead, ThermoVault obtained consent from the social housing management to adapt the invitation letter due to the non-invasive, minimal-risk intervention, which we deemed sufficient.

**Intervention design**

Using RCTs in both study settings, we assigned tenants to a control group that received a classic letter, or intervention group with a letter with additional information on descriptive social norms and related experienced benefits from prior uptake of the technology from tenants of the same social housing company. In both studies, we used information based on real-world data from test rollouts two years prior.

---

[2] This information is based on data of prior rollouts of ThermoVault in social housing in Belgium.

The classic letter contained information on ThermoVault, the technology itself with general description and a meeting appointment with contact details. For the intervention, a graphic and a small abstract were added, informing tenants about the number of households in the same social housing company already using the technology, and their benefits from it. Figure 1 illustrates the intervention, which was included on the first page of the letter.[3]

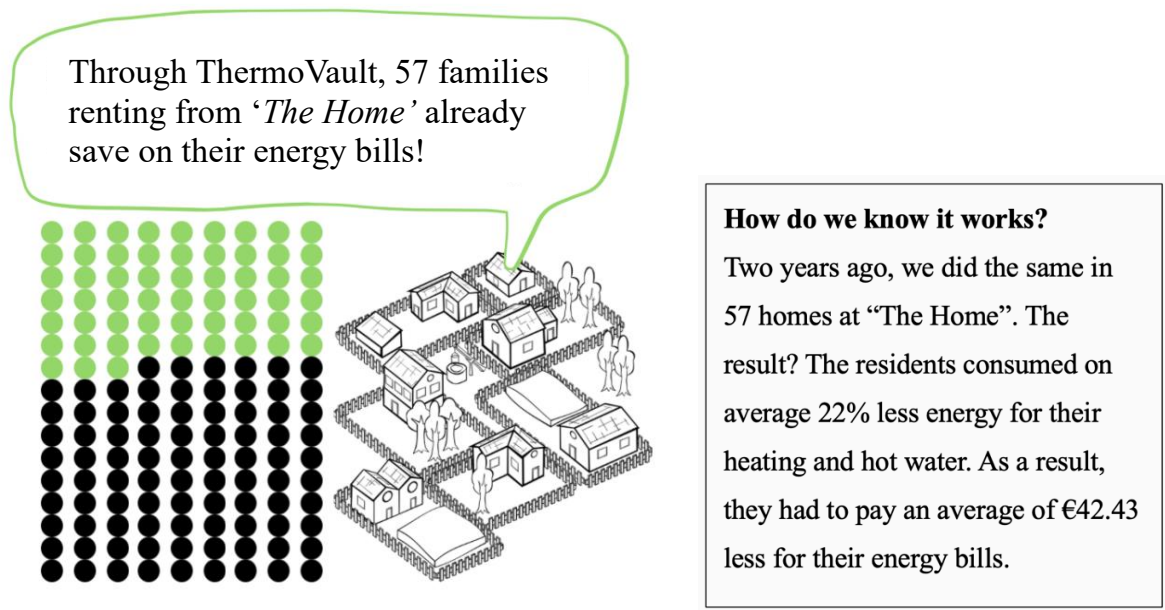

Figure 1. Intervention design (added to letter, see Appendix for context).

In the first study, randomization was only possible on the apartment block level, i.e. apartment blocks were randomly assigned to either the control or intervention group. The second study allowed for randomization on apartment level, so apartments within apartment blocks were randomly chosen to receive either the control or intervention letter. Sample sizes were determined through households priorly selected for the technology rollout, which is why we decided to stick with the minimal two-group design in both studies.

---

[3] Note that the original information in the letter was displayed in Dutch, and was translated (and back-translated to English) by native speakers to ensure proper wording. For privacy reasons, we changed the name of the social housing company in the Figure.

**Data collection.**

The dependent variable in our setting is the actual uptake of the retrofit technology, i.e., a successful installation in a contacted household. This data depends on the implementation of the technical rollout realized by ThermoVault. Each apartment received in their letter an appointment for installation, which tenants could either accept by default or change by contacting ThermoVault staff. If tenants were not at home, a second attempt was made on one of the subsequent days. For all apartments included in the trial, we received an overview on whether installation was successful in the end, and if not, for what reason the installation was not deployed. In both studies, the dependent variable was coded 1 if the installation was successful, and 0 if the tenant declined the installation or did not open the door throughout the whole trial period.

## Study 1

**Sample Details**

In the first study, on 01.02.2022, letters were sent by the social housing management to N = 187 apartments, with apartment blocks (N = 7) randomized to letter groups. N = 141 apartments in four apartment blocks received the control letter, and N = 46 apartments in three apartment blocks received the intervention letter.[4]

---

[4] The small sample size in the trust intervention group was partially caused by postal issues, where some letters were not sent to priorly defined apartments, which then had to be excluded.

**Results**

Out of the 187 apartments, 157 installations were successful, a 84% success rate. In the control group (N = 141), 113 apartments accepted the installation, while in the intervention group (N = 46), 44 installations were successful. Figure 2 shows that for installation success, the intervention letter group has a higher mean (*M* = .96. *SD* = .21) than the control group (*M* = .80, *SD* = .40).

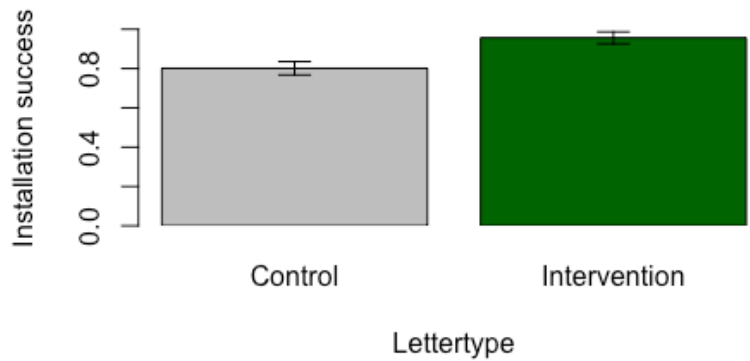

*Figure 2. Installation success rates (0-1) by letter type, study 1. Error bars are 95% confidence intervals.*

We performed a logistic regression to ascertain the effect of letter group on the likelihood that installation was successful, and report standardized estimates. We find that group was a significant predictor (*ß* = 1.7, *p* = .024, 95% CI [0.44, 3.55]) for installation success, with McFadden's $R^2$ = 0.046. Compared to the control condition, the odds of installation success increased by OR = 5.45 (95% CI [1.25, 23.85]) in the intervention group.

# Study 2

**Sample details**

To replicate study 1, and to improve methodological constraints such as randomization procedure and the differences in sample sizes between groups, we conducted a second study: letters were sent to 177 apartments in three buildings of a social housing company in Belgium. We randomized apartments into the two groups (using R), with no special assignment regarding apartment block. Letters were sent by the social housing company on 27.09.2022 to $N_{Control} = 88$ apartments, and to $N_{Intervention} = 89$ apartments. We had to exclude apartments where installation was technically not feasible, as in these cases, we can't draw a clear conclusion if tenants would have accepted the technology or not. This was true for $N = 49$, which were removed before final analysis. This led to a final sample of $N = 128$ in study 2 ($N_{Control} = 66$, $N_{Intervention} = 62$).

**Results**

Out of the 128 apartments, 99 installations were completed, a 77% success rate. In the control group ($N = 66$), 48 apartments accepted the installation, while in the intervention group ($N = 62$), 51 installations were successful. Figure 3 shows that for installation success, the intervention letter group has a slightly higher mean ($M = .82. SD = .38$) than the control group ($M = .72, SD = .45$).

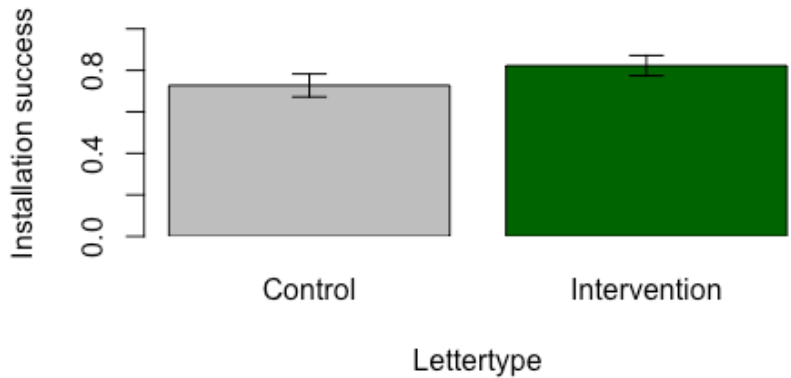

*Figure 3. Installation success rates (0-1) by letter type, study 2. Error bars are 95% confidence intervals.*

In a first step, we did not find a significant difference between groups, with intervention group not being a significant predictor for success ($ß = .55$, $p = .20$, 95% CI [-0.28, 1.43]). Considering model fit, we found that the model did not show a good fit to our data (adjusted McFadden's $R^2 = .0122$). Exploring our data further, we found a pattern which points to the relevance of considering the building block level, depicted in Figure 4.

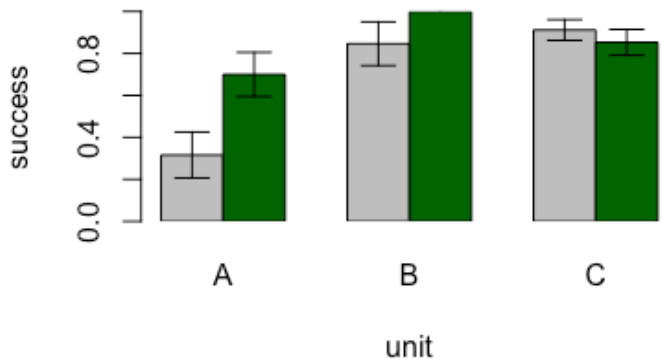

*Figure 4. Comparison between groups in success, by apartment blocks (unit). Error bars are 95% confidence intervals. Gray = control; green = intervention. Note: The absence of the error bar in Unit-B is due to the acceptance rate in this group being 100% (i.e. success = 1), leading to no variability within that group.*

We performed a logistic regression to ascertain the effect of letter group on the likelihood that installation was successful when taking building block into account in the model. This increased our model fit substantially (adjusted McFadden's $R^2$ = .211). We found a significant main effect of the intervention ($β$ = 1.62, $p$ = .020, 95% CI [0.31, 3.05]) while controlling for the interaction with the apartment block, with Block A significantly different from apartment Block C; ($β$ = -2.20, $p$ = .035, 95% CI [-4.33, -0.20]). In this model, compared to the control condition, the odds of installation success increased by OR = 5.06 (95% CI [1.30, 19.70]) in the intervention group for Block A.

## Discussion

Findings from two randomized controlled trials conducted in Belgian social housing suggest that our intervention, building on social norms and concrete experienced benefits, is promising for increasing adoption of retrofit technologies among social housing tenants. In both studies, the addition of a small graphical and textual norm and benefits information about the adoption rate of other households boosted the willingness to have the technology installed. In the first study, the odds for installation success in the intervention group increased by over 5 times, which we replicated in a second study for one of the apartment blocks. While the financial benefits alone (free device, with generated savings in energy costs) had not been enough to convince all households before, communicating benefits through descriptive social norms constituted an additional motivation for sustainable technology adoption in this context.

The underlying mechanisms of the positive effect of the intervention could be a decrease in uncertainty about the behavior due to the heuristic value of other people's decision (Constantino et al., 2022). Reduced uncertainty might also be related to an increase in trust, which affects the evaluation of a technology's risks and benefits (Huijits et al., 2012). In former studies, only if the trust value of a particular norm has been high, it has increased the likelihood of adopting that norm (Itaiwi et al., 2018). Thus, conveying benefits via descriptive social norm information likely enhanced trust and diminished risk perceptions related to the adoption behavior (Veflen et al., 2020; Zou & Savani, 2019). A direct measure of trust as the explanatory mechanism between the appraisal of our interventions and the outcome of adoption could be studied in the future.

Additionally, future research should investigate moderators of social norms, like the extent of individuals' identification with the reference group, to better understand the intervention's mechanisms (Cialdini & Jacobson, 2021). The level of social identification might explain the differing pattern in the control group between different housing blocks in the second study (with B and C having much higher baseline values in control than A). Further, it would be interesting to investigate concepts as collective efficacy or social identification (Fritsche et al., 2018), which seem of particular import in the context of social housing communities and social integration challenges (Quilgars & Pleace, 2016).

There are several limitations to acknowledge in this study. Firstly, no demographic information about the sample was accessible to be included as a moderator. However, access to social housing in Belgium is governed by a specific criteria set[5] and legal guidelines dictate size and pricing of social housing, with rents calculated based on income, and decreasing with lower income (VMSW, 2022). We therefore can assume a quite homogenous sample in our studies. As we aimed for experimental evidence particularly in social housing settings, a highly under-researched sample (McCabe et al., 2018), generalizations beyond this sample should, if at all, be made with care; and, especially due to our small sample sizes and uneven distribution in study 1, our results should be interpreted with caution. In the second study, the significant effect was only observed in one of the apartment blocks, which appears to be due to a ceiling effect in the other two blocks; still, a careful interpretation is in order, as other factors might be at play that might limit generalizability of the findings. Finally, we operationalized participants who were not at home as rejecting the technology, as they were informed about time of install and approached for it multiple times; however, we have no insights whether their rejection was due to a lack of interest or other reasons.

---

[5] https://www.vlaanderen.be/sociaal-woonbeleid/verhuren

    Despite its limitations, this research contributes to the growing body of literature on the efficacy of interventions for promoting the adoption of energy-efficiency technologies, with a specific contribution related to the social housing context. The findings from two RCTs conducted in Belgian social housing settings suggest that a simple communication intervention with trustworthy information about successful adoption of other social housing tenants can be a promising intervention for increasing the adoption of retrofit technologies. Future research should explore interventions' potential to achieve the European Union's sustainability goals in the energy sector, particularly in specific contexts like social housing.

# Appendix A

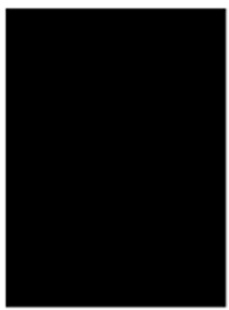
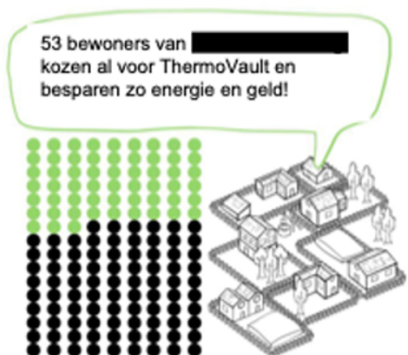
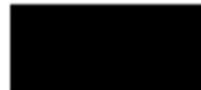
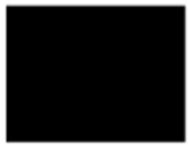
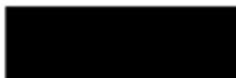
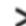

*Figure 5. Letter with intervention.*

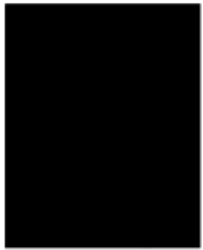

«Voornaam» «Familienaam»
«Straat_eenheid» «Huisnr_eenheid»
«Busnr_eenheid»
«Postnr_eenheid» «Gemeente_eenheid»

2022

Onze referentie
Contactpersoon
Onderwerp  Slim besparen op de energiekosten van jouw woning

## Beste «Voornaam»,

Energie wordt alsmaar duurder. Daarom werken we bij De Ideale Woning samen met de firma **ThermoVault** om jouw woning te voorzien van een **energiebesparend systeem**. We plaatsen een slim apparaat in je elektrische **boiler** dat ervoor zorgt dat die ongeveer 20% minder energie verbruikt. De Ideale Woning biedt het systeem **gratis** aan. Het werkt **automatisch** en je kan alle toestellen blijven gebruiken zoals voorheen.

### Wat zal er gebeuren en wanneer?
We stelden en erkende installateur, 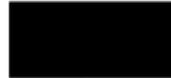, aan voor het plaatsen van dit energiebesparend systeem. Het gaat om een klein toestel dat wordt geplaatst in de **elektrische boiler, verwarmingstoestel en in de zekeringkast**. De installatie duurt ongeveer een uur, zonder breek- of slijpwerk. Achteraf kan je meteen je verwarming en warm water weer gebruiken.

**Binnenkort** krijg je van ons nog een **brief** waarin we een **afspraak** voorstellen waarop de installateur langskomt. Je kan die afspraak dan nog **verplaatsen** als ons voorstel niet past.

### Wat verwachten we van jou?
- Maak op voorhand de **ruimte** rond de zekeringkast, de elektrische boiler en je verwarmingstoestel **vrij**.
- Hou een recente **elektriciteitsfactuur** klaar, daar staan alle technische gegevens op die de installateur nodig heeft.
- Zorg ervoor dat er iemand **thuis** is op het moment van de afspraak om de installateur binnen te laten.
- Als de installateur klaar is, zal hij je vragen om iets te **ondertekenen**. Het is **belangrijk** dat je dat doet, want we hebben je **toestemming** nodig zodat het apparaat je

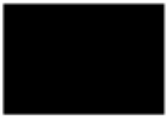 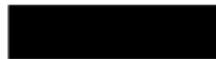

>

*Figure 6. Letter (control) without intervention*

## Appendix B: Social norms & energy conservation

Social norms operate as powerful tools in shaping behavior, especially in the context of environmental conservation. Feedback comparing personal energy saving with others' performance triggers upward or downward social comparison, influencing individuals' actions towards more sustainable practices (Cialdini & Jacobson, 2021). Several field studies find that communicating descriptive and/or injunctive norms has significant impact on energy related practices in public (Bator et al., 2014; Bergquist & Nilsson, 2016; Dwyer et al., 2015; Liu et al., 2016) and residential settings (Allcott, 2011; Ayres et al., 2013; Bhanot, 2021; Schultz et al., 2007, 2015). Table 1 gives an exemplary overview on these studies, demonstrating positive effect of social norms on energy conservation, whether descriptive and/or injunctive social norms were and which kind of behavior was targeted.

| Study | Behavior | | Norms |
|---|---|---|---|
| Bator et al. 2014 | Public | Turning off computers/monitors in campus laboratory | Descriptive & injunctive norms |
| Bergquist & Nilsson 2016 | | Energy conservation behavior in public bathrooms | Descriptive & injunctive norms |
| Dwyer et al. 2015; | | energy conservation behavior in public bathrooms | Descriptive norms |
| Liu et al., 2016 | | petition signing addressing energy consumption in a University campus building | Descriptive norms |
| Schultz, 2007 | Household | Energy conservation | Descriptive & injunctive norms |
| Allcott , 2011 | | Electricity usage | Descriptive norms |
| Bhanot, 2021 | | Water conservation | Injunctive norms |

| Ayres et al., 2013 | | home electricity and natural gas usage | Descriptive norms |
| Schultz et al., 2015 | | Electricity consumption | Descriptive norms |

*Note: This overview is not meant to be exhaustive, but rather to give some examples on field research showing positive effects of social norms for energy conservation.

Research finds that aligning descriptive and injunctive norms in communication enhances their effectiveness in promoting energy-saving behaviors (Bonan et al., 2020) and can prevent potential boomerang effects of descriptive norms (Schultz et al., 2007). Moreover, the effectiveness of social norm interventions varies based on factors such as personal identification with the reference group and the proximity of the norm to the individual's current behavior (Abrahamse & Steg, 2013). The influence of norms is stronger when displayed by others with closer levels of a behavior (Bergquist & Nielsson, 2018), and norm-based messages are more persuasive when they highlight in-group identification (e.g. Lede et al., 2019), demonstrating social identification as crucial moderator of social norms (Cialdini & Jacobson, 2021). The effect of norms on energy conservation behavior and energy related practices was considered in multiple meta-analyses and reviews, summarized shortly in Table 2 below.

| **Source** | **Energy behavior** | **#studies included (related to energy behavior)** | **Results & overall conclusions** |
|---|---|---|---|
| Delmas et al., 2013 | Energy conservation | 12 papers (including 37 observations) | Weighted average treatment effects = 11.5% reduction in energy conservation followed by social comparisons (i.e. norms) |

| Farrow et al. (2017) | Energy use | 11 papers | 61% of papers for energy use find supportive evidence, Although social norm interventions effective, still knowledge gaps |
| --- | --- | --- | --- |
| Cialdini & Jacobson (2021 | Energy conservation | 14 papers | 13 out of 14 demonstrated supportive evidence of social norms on energy conservation behavior |
| Composto & Weber | (1) Reduction of household energy demand (overall); (2) electricity use ; (3) Investment in energy efficiency | 158 papers (overall, 1) including 123 (for 2); 17 (for 3) | Mixed results for effectiveness of norm interventions; small to medium effect for communicating a norm on energy savings; small but robust effect in the adoption of new technologies |